# Testing the Effect of Hydrophilic and Hydrophobic Coatings on the Speed of a Ball through Water

Natalie Wiegand
and Philippa Thomas

Science Department
St. Matthew's Parish School
1370 Bienvenieda Avenue
Pacific Palisades, California 90272

**Abstract:**

Data is presented that confirms that hydrophobic coatings reduce friction between objects and water. The results show that the average time it took for the ball with the hydrophobic coating traveled at an average of about 6 inches per second. The ball with the hydrophilic coating traveled at a slower pace, moving at an average of about 5 inches per second and the ball without a coating traveled at an average of about 4.8 inches per second.

## I. Introduction

Water can interact with surfaces differently depending on whether the surface is hydrophobic or hydrophilic. Hydrophobic means that the substance does not combine or mix well with water. Athletic apparel, such as Nike Dry-Fit, uses hydrophobic coating on the material. This wicks the moisture from the fabric, keeping you dry in all weather. Some examples of day to day hydrophobic substances are oil and Crisco. Hydrophilic means that the substance mixes with water because it has strong polar groups that readily interact with water. Soap and sugar are also examples of hydrophilic substances and dissolve well. This interesting property has led to numerous experiments and led us to conduct ours.

The purpose of this experiment is to test whether coating objects with either a hydrophilic or a hydrophobic coating will cause the object to move through water more quickly. This was done by coating a golf ball in a hydrophilic coating, a hydrophobic coating and one without coating and dropping them into a tube filled with water. The time it took for the ball to pass from one point in the water to the next point in the water was recorded. It was concluded that the hydrophobic coating reduced friction between the ball and the water, thereby permitting it to pass through the tube significantly faster (Figure 1).

## II. Method

To begin the experiment, first set up the apparatus (See figure 1). To seal the bottom of the six foot tube, cover one end of the tube in duct tape. Then, affix duct tape going in the opposite direction on top of the previous layer. To finalize the seal, use a standard caulking gun and coat the duct tape in caulk. Wait at least two days or until the caulk fully dries to fill it with water.

To coat the balls with the hydrophilic coating moisten a cotton swab with hydrophilic coating, we used isopropyl alcohol, and completely cover the surface of the ball with coating using the cotton swab. To set the coating, steadily heat the ball with a heating mechanism and rotate until coating is set. For the hydrophobic coating, make sure that the surface is clean and dry. If the ball is dirty, wash with a mild detergent, then rinse well with water. When applying WX2100, the temperature should be between 45 and 95 degrees Fahrenheit. Next, shake the can vigorously for one to two minutes before application to the ball. Then, spray, holding the can vertically eight to ten inches from the surface of the ball and evenly cover. Shake the can every 15 seconds during the application. The coating should appear frosty as it begins to dry. Allow at least two hours before exposure to water. *Slightly modified from WX2100 Coating for the hydrophobic coating instructions.

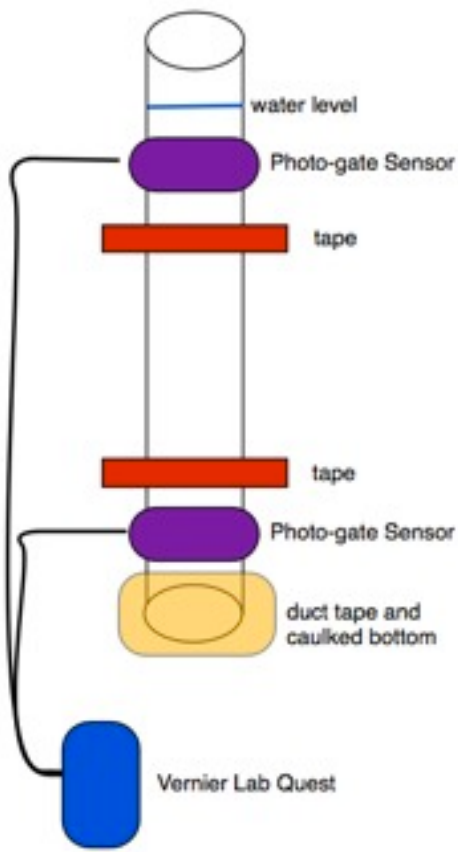

**Figure 1.**  The set-up of the apparatus includes a clear six foot pipe, two photo-gate sensors and a Vernier Lab Quest system.

To perform the experiment, fill a six foot clear polycarbonate tube to the 5' 10" mark with tap water (Figure 1). Then, fix it vertically to a wall. Then, place two Vernier Lab Quest photogate sensors were at the top and bottom of the tube to record the time it will take for the ball to pass from one sensor to the next. Then, start the photogate program on the Vernier Lab Quest system. Drop the ball starting with the ball with a hydrophobic coating from slightly above the top of the tube and let it fall through to the bottom. When the Vernier Lab Quest has finished recording the time, extract the ball from the tube. Then, repeat using the ball with a hydrophilic coating and using the ball with no coating.

## III. Results

We found from our results that when using hydrophobic coating, the ball falls through the tube faster. Also, we found that when the ball was left uncoated, the average time it took to travel through the tube was very similar to that of hydrophilic coating. We found that the average time it took the ball to go through the tube was 11.8 seconds ($\approx$6 in/sec.) when it was coated with hydrophobic coating. While when using hydrophilic coating the ball traveled at a slower rate, taking an average of about 13.5 seconds ($\approx$5 in/sec) to pass through the tube. When we used no coating it took an average of about 13.8 seconds ($\approx$4.8 in/sec.). The hydrophobic coating permitted less drag on the ball, enabling it to move a little more than 7% faster than the ball without any coating. Through this, we proved that some objects underwater can move quicker when coated with hydrophobic coating and hydrophilic coating is not very effective in increasing the speed though it is faster than having no coating.

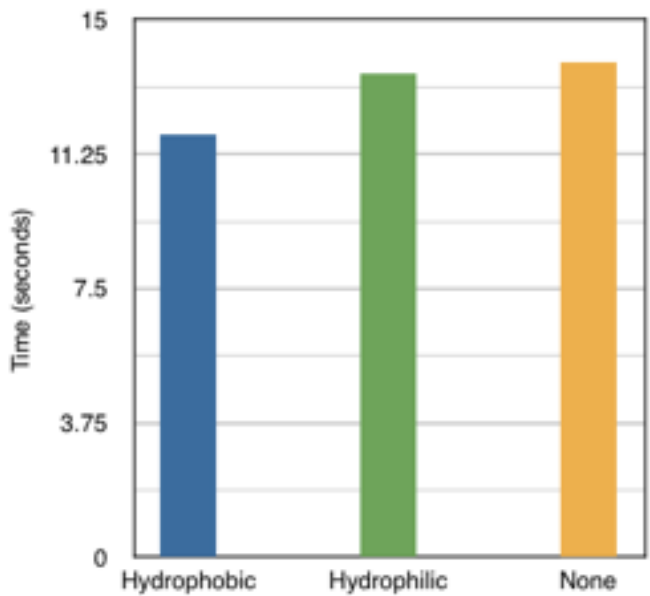

**Figure 2.** We show here a plot of the time it took for the both of the treated balls and the un-treated ball to fall through a column of water 180 cm tall.

| Coatings | Trial 1 | Trial 2 | Trial 3 | Trial 4 | Trial 5 | Trial 6 | Trial 7 | Trial 8 | Trial 9 |
|---|---|---|---|---|---|---|---|---|---|
| Hydrophobic | 12.309 | 12.748 | 11.641 | 12.305 | 11.271 | 11.237 | 11.814 | 11.362 | 11.378 |
| Hydrophilic | 13.150 | 13.438 | 13.709 | ~~15.306~~ 13.656 | 13.300 | ~~12.221~~ 13.456 | 13.749 | 13.415 | 13.679 |
| None | 13.275 | 14.476 | 14.024 | 14.646 | 13.063 | 13.385 | 13.208 | 14.054 | 14.285 |

**Table 1.** We show here the results acquired from our experiment for both of the treated balls and the un-treated ball to fall through a column of water 180 cm tall. This includes both original results and results that were re-done after deciding that they were outliers.

### IV.   Analysis

Our data indicates that hydrophobic coating is significantly affecting the ball, permitting it to travel faster through the water than the hydrophilic coating. Using this information, one could choose to coat boats, submarines and other underwater vehicles so that their speed would increase. Also, this information can explain why certain underwater animals have a hydrophobic skin. They most likely evolved to have this trait because having hydrophobic skin proved to be an advantage over other types of skin. Also, some insects that do not swim in the water but move on the surface of the water have hydrophobic legs. This keeps their legs dry and might allow for less drag from their legs on the water, making them go faster. Overall, we learned that the coating on something can greatly affect its speed through the water.

To show how beneficial to the environment using hydrophobic coating would be to the fuel efficiency of a cruise ship here are some statistics:
The average cruise ship uses 200 gallons of gas per nautical mile traveled. If boat manufacturers were to coat boats with hydrophobic coating one could save 14 gallons of gas per mile traveled. The average cruise ship travels 625 nautical miles per day. This means the average cruise ship uses 125,000 gallons of gas per day. If this ship was coated with hydrophobic coating it would save about 8,750 gallons of gas. While this does not seem like a great amount of fuel saved, if all cruise ships were coated with hydrophobic coating, it would greatly increase the fuel efficiency of the boats, according to our data.

Another example of how beneficial the use of hydrophobic coating is to not only the Earth, but people, is in municipal water mains. Every year, cities spend a great amount of vital energy pumping water through the local pipes. If the Department of Water and Power were to coat the inside of pipes with a hydrophobic coating, the water would slide easier through the pipes, using much less energy per year and saving taxpayers money.

We took various steps to eliminate error in our research though encountered error none the less. Because we wanted to eliminate as much human error as possible, we decided to use the Vernier Labquest Photogate sensors to record the amount of time it took the ball to fall through the tube. Although we eliminated the form of human timing error, we still had many other sources of error. The first is the variability of the drop height. We tried to drop the balls from the same height each trial, but because we did not have a line or point where we were consistently dropping the ball from, it changed slightly, which caused a few outliers. Another source of error we could have had was the ball hitting the sides of the tube while dropping to the bottom. This would cause the ball to slow down, causing more error.

When looking at our results, we eliminated the results that were clearly outliers (Figure 3). We are not sure what went wrong in our experiment to cause the outliers, but we know something went wrong because the results were completely out of the range of the other data. To make up for the data we lost, we did new trials that turned out close to our average. These extra trials allowed us to get even more accurate data.

From this experience there are a couple of elements that could be changed. Throughout the experiment process, we discovered that many of the key pieces to our experiment needed to be altered or completely taken out. First, we found that when the ball was coated in hydrophobic coating, and placed in a pantyhose bag that was coated in hydrophobic coating, it floated. We also tried to drop a ball with no coating in a pantyhose bag; it also floated. We then came to the conclusion that the ball would not sink with the addition of the pantyhose bag unless weights were added. If we would have known this before we started the experiment, we would have gone straight to not using the bag and simply coating the balls. Secondly, it would have been beneficial to have used our time more efficiently toward the beginning of the process. We spent a great amount of time simply figuring out how to get our large pipe to stand up. We first tried to position our pipe in a bucket full of concrete. This idea did not turn out well because the pipe did not stick and stay in the concrete, and we would have been stuck with carting around a 100 pound bucket. After we had decided against the concrete bucket method, we realized we could simply duct tape the pipe to a wall. If we had figured out that we could stand the pipe up by attaching it to a wall from the beginning, we would have had more time for the completion of our experiment. In conclusion, if we did our experiment again, we would have chosen a heavier ball and we would have found a good way to attach our pipe earlier in the process.

There are many aspects of this experiment that could be further experimented with. Some elements that could be added are different shaped objects. With this addition, we could have tested for a difference with the different shapes and coatings. This would combine the element of hydrodynamics to our experiment. Another possible addition would be the incorporation of a mini submarine or a boat that we could coat with the different coatings. This would allow for more realistic results for the use in marine travel. To make an even more realistic marine environment, the balls, submarine or boat could be tested in salt water. Another possible adaptation to the experiment would be to test hydrophobic and hydrophilic coatings with different liquids and liquids with different pH and salt levels to see if it affects the coating.

## V.  Conclusion

In this paper we have presented data confirming that hydrophobic coatings allow certain objects in tap water to travel more quickly through the water than without a coating. Furthermore, hydrophilic coatings on specific objects allow them to pass through tap water slightly faster than without. Our results show that the average time it took for the ball with the hydrophobic coating traveled at an average of about 6 inches per second. The ball with the hydrophilic coating traveled at a slower pace, moving at an average of about 5 inches per second and the ball without a coating traveled at an average of about 4.8 inches per second.

With this information, hydrophobic coating could be use to make water travel more effective as well as making instances where water interacts with surfaces more energy efficient.

## Acknowledgments

We are grateful for Tadd T. Truscott for sharing his knowledge and experience with our topic. We are grateful for Admiral Robert Janes for sharing his knowledge about Naval Vessels and marine travel. We are grateful for Michigan Institute of Technology and Brigham Young University for sharing advice on researching this topic. We are grateful for St. Matthew's Parish School for supporting this experiment. We are grateful for Mr. J. Brownridge for his suggestions concerning our paper.

## References

Miller-Keane Encyclopedia and Dictionary of Medicine, Nursing, and Allied Health, Seventh Edition. 2003 © by Saunders, an imprint of Elsevier, Inc.

Truscott, Tadd T. "Tadd Truscott." *Tadd Truscott*. N.p., n.d. Web. 18 Mar. 2012.

"Sleek Swimsuits, Inspired by Penguins." *Specialty Fabrics Review*. Specialty Fabrics Review, May 2011. Web. 14 May 2012.

"Are Hydrophobic Boat Hulls Faster in Water?" *Slippery Sailboats*. South Hill Enterprise. Web. 14 May 2012.

"How Much Fuel Does a Cruise Ship Use?" *How Much Fuel Does a Cruise Ship Use?*. Web. 14 May 2012.

MacKenzie, Angus. "Ultra-Ever Dry Hydrophobic Coating." *Ultra-Ever Dry Hydrophobic Coating*. Ultra-Tech, 24 Feb. 2013. Web. 05 Mar. 2013.

Lama, Dilraj, Vivek Modi, and Ramasubbu Sankararamakrishnan. "Behavior of Solvent-Exposed Hydrophobic Groove in the Anti-Apoptotic Bcl-XL Protein: Clues for Its Ability to Bind Diverse BH3 Ligands from MD Simulations." *PLOS ONE*. PLOS ONE, 28 Feb. 2013. Web. 05 Mar. 2013.